# Hydrogel modified evaporation interface for highly stable membrane distillation

*Yanni Ma†, Zehua Yu†, Xifan Fu, Zhi Huang\*, Tenghui Qiu, Na Zhao, Huidong Liu, Kang Liu\**

†These authors contributed equally to this work.
Laboratory of Hydraulic Machinery Transients, School of Power and Mechanical Engineering, Wuhan University, Wuhan, Hubei 430072, China.
\* Corresponding to: kang.liu@whu.edu.cn; zhihuang@whu.edu.cn

**Abstract**
Surface effect of low-surface-tension contaminants accumulating at the evaporation surface can easily induce membrane wetting in the application of membrane distillation, especially in hypersaline scenarios. In this work, we propose a novel strategy to eliminate the surface effect and redistribute contaminants at the evaporation interface with simply incorporating a layer of hydrogel. The as-fabricated composite membrane exhibits remarkable stability, even when exposed to extreme conditions, such as a salt concentration of 5M and surfactant concentration of 8 mM. The breakthrough pressure of the membrane is as high as 20 bars in the presence of surfactants, surpassing commercial hydrophobic membranes by one to two magnitudes. Combined study of density functional theory and molecular dynamics simulations reveals the important role of hydrogel-surfactant interaction in suppressing the surface effect. As a proof of concept, we also demonstrate the stable performance of the membrane in processing synthetic wastewater containing surfactants of 144 mg $L^{-1}$, mineral oils of 1g $L^{-1}$ and NaCl of 192 g $L^{-1}$, showing potential of the membrane in addressing challenges of hypersaline water treatment and zero liquid discharge processes.

## 1. Introduction

Desalination of seawater and inland brackish water emerges as a promising solution to combat freshwater scarcity.[1] However, limited by the low recovery ratio of the current desalination technologies, desalination plants produce substantial quantities of concentrated brine, the disposal of which poses significant environmental challenges, particularly concerning land vegetation and the aquatic ecosystem.[2] Thus, there has been a growing imperative towards maximizing reuse of liquid waste to achieve zero liquid discharge.[3] Zero liquid discharge necessitates the continuous concentration of water to a hypersaline state, which heightened demands on both materials and systems to effectively resist wetting and fouling issues.[4] Membrane distillation—a thermally driven separation process that only allows the vapor molecules to pass through a microporous hydrophobic membrane—possesses high tolerance of salinity and near-complete rejection of nonvolatile solutes, rendering it highly promising for achieving high flux zero-liquid discharge.[5] In traditional membrane distillation, Laplace pressure of the meniscus at the liquid/vapor interface prevents liquid infiltration into the hydrophobic membrane (**Figure 1**a), and the magnitude of this Laplace pressure is determined by the pore size and surface tension of the liquid.[6] The surface tension is sensitive to inclusions within the water.[7] For hypersaline water with complex components, low-surface-tension contaminants naturally accumulates and boosts the concentration of contaminants at the liquid-vapor interface.[8] This surface effect can easily induce membrane wetting. High salinity further exacerbates the surface effect because the presence of NaCl increases the affinity of low-surface-tension contaminants and accelerates the adsorption equilibrium process at the gas-liquid interface.[9] To avoid the contaminants accumulation, a hydrophilic layer with small pore size has been proposed to block the contaminants by size-seizing effect, forming a Janus membrane.[10] However, small-size solutes or organics easily get stuck in the pores of the hydrophilic layer. The strategy of blocking inevitably renders the membrane susceptible to issues such as scaling or fouling, particularly in hypersaline wastewater.[10b, 11] Hydrogels are three dimensional cross-linked networks of hydrophilic polymer chains with water occupying the interstitial spaces.[12] These polymer chains, despite being a relatively small component in terms of volume, provide hydrogels with abundant functional groups network that exhibit strong interactions with low-surface-tension contaminants.[13] The invisible network may offer opportunity to manipulate the surface effect of contaminants accumulation and redistribute the contaminants at the evaporation interface without blocking. Based on this idea, we propose a strategy to modify the evaporation interface to counteract wetting issues in hypersaline solution by the incorporation of a hydrogel layer onto a hydrophobic polytetrafluoroethylene (PTFE) membrane (Figure 1b). This hydrogel modified PTFE membrane possesses a breakthrough pressure of 20 bars, even with the existence of surfactants, and shows stable membrane distillation performance, even when exposed to solutions with a salt concentration of 5M and a surfactant concentration of 8 mM, — challenges that have not been previously addressed in the existing literature. Density functional theory (DFT) calculations and molecular dynamic (MD) simulations were conducted to explore the microscopic mechanism underpinning the anti-wetting properties.

Finally, we demonstrate the stable anti-wetting and anti-fouling performance of the membrane in hypersaline synthetic wastewater with surfactants and mineral oil, showing its potential for achieving zero liquid discharge.

## 2. Results

### 2.1. Membrane desalination with hydrogel modified PTFE film

To demonstrate the concept of the hydrogel modified hydrophobic membrane, we fabricated a composite membrane by copolymerizing the polypropylene (PP) side of a commercial PTFE membrane with a sodium polyacrylate (PSA) hydrogel layer (Figure S1a, Supporting Information). The PTFE membrane possesses a pore size of ~100 nm (Figure S1b, Supporting Information). The obtained hydrogel-modified PTFE (H-PTFE) membrane exhibits good flexibility (Figure 1c). The hydrogel coating has a thickness of 150 μm, penetrating pores of the microporous PP layer (Figure 1d). Existence of ether bonds demonstrates the successful grafting between the PP and hydrogel layer (Figure S1c, Supporting Information). The partial penetration and ether bond formations anchor the hydrogel layer firmly onto the PTFE membrane. The hydrogel side of the H-PTFE membrane is superhydrophilic with a water contact angle near zero degree and exhibits excellent oleophobicity underwater. The PTFE side is hydrophobic and underwater oleophilic (Figure S2, Supporting Information).

To evaluate the performance of the membrane, we tested it in membrane distillation using brine feeds with varying salt and sodium dodecyl sulfate (SDS) concentrations (Figure S3, Supporting Information). SDS is a typical surfactant in industry, and serves as a representative low-surface-tension contaminant. As shown in Figure 1e, F, the hydrogel modified membrane is highly stable with a constant mass flux and 100% salt rejection rate in all tested solutions, even in a near-saturated NaCl concentration of 5 M and SDS concentration near critical micelle point of 8 mM. Under the high salt and SDS concentration, we further increase the applied static pressure at the feed side, and the membrane continues to perform reliably under an external pressure of 3.2 bar (Figure 1g). As a comparison, commercial PTFE membrane is wetted under the salt concentration of 0.6 M and SDS concentration of 0.2 mM. As salt concentration increases, the membrane's tolerance for SDS decreases (Figure 1e). Compare the performance with previous studies involving omniphobic membranes (OMs) and Janus membrane (JMs),[14] the hydrogel modified membrane shows much superior capacity in hypersaline and surfactant-rich environment (Figure 1h).

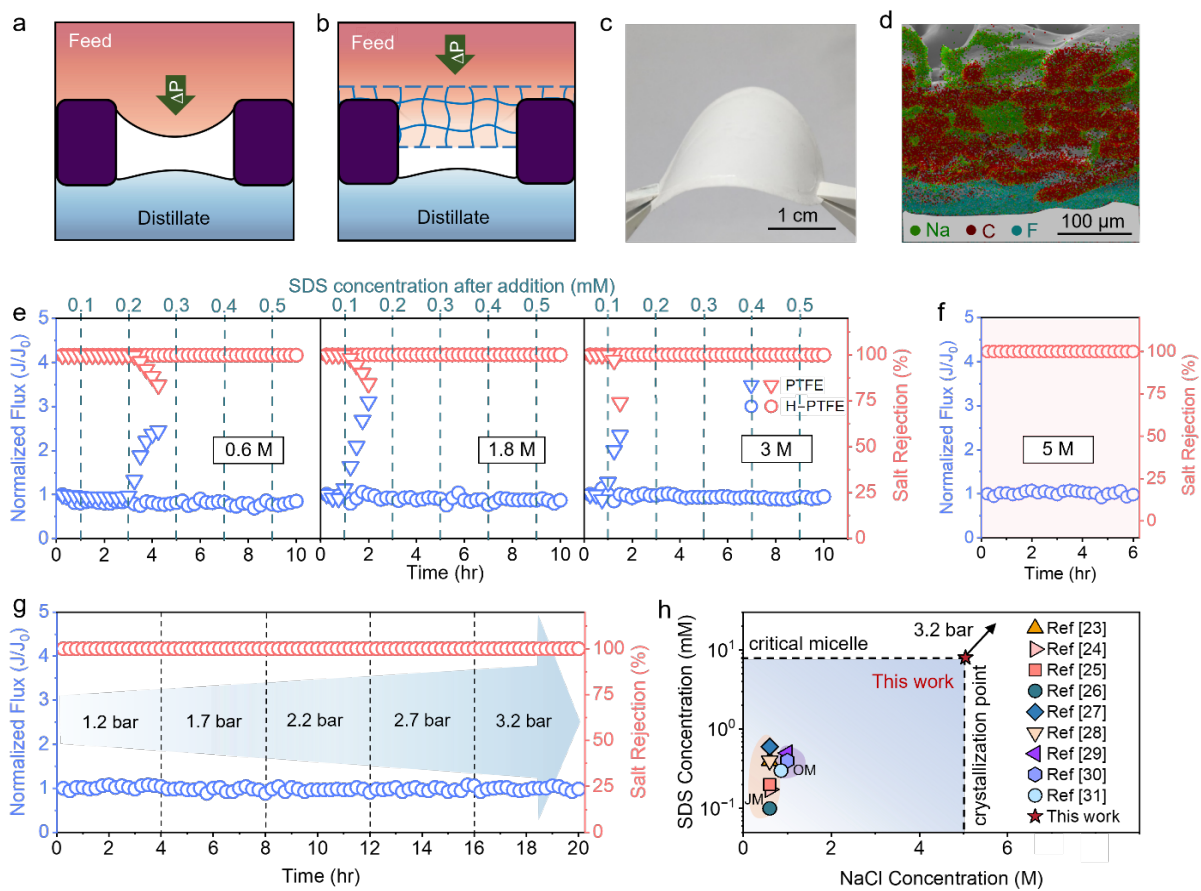

**Figure 1.** Concept of hydrogel modified evaporation interface and its anti-wetting performance. a-b), Schematic diagram of traditional membrane distillation and hydrogel modified membrane distillation process. c), Photographic image of hydrogel modified PTFE membrane. d), SEM cross-sectional view of the H-PTFE membrane. The energy dispersive spectrum (EDS) was used to distinguish the composition of the membrane. The green, red and blue region denote the PSA, PP and PTFE, respectively. e), Normalized vapor fluxes (blue points) and salt rejection rates (red points) for PTFE and H-PTFE membrane in direct contact membrane distillation (DCMD) operation at different NaCl concentrations from 0.6 to 3 M. The green dashed lines denote the addition of SDS. The initial vapor fluxes of the PTFE membranes at salt concentrations of 0.6, 1.8 and 3 M were 18.44, 16.2 and 15.22 L m$^{-2}$ h$^{-1}$, respectively. The fluxes of the H-PTFE were 12.62, 10.31 and 9.24 L m$^{-2}$ h$^{-1}$, respectively. f), Distillation performance of H-PTFE membrane in the feed with 5 M NaCl and 8 mM SDS. The initial vapor flux was 6.62 L m$^{-2}$ h$^{-1}$. g), Distillation performance of H-PTFE membrane at an increased feed hydraulic pressure in the feed with 5 M NaCl and 8 mM SDS. h), Comparison of feed salt and SDS concentration of membrane distillation with the literature data working with omniphobic membranes (OMs) and Janus membranes (JMs) (Table S1).

The upper dotted line represents the critical micelle concentration of SDS, and the right dotted line represents the crystallization point of NaCl.

Breakthrough pressure, also known as the liquid entry pressure, was tested to quantitatively characterize the wetting performance of the membrane (Figure S4, Supporting Information). As shown in **Figure 2**a, breakthrough pressure of the unmodified PTFE membrane is 2.7 bar. With an increase in SDS concentration, the breakthrough pressure decreases, approaching 0.4 bar at an SDS concentration of 0.5 M. In contrast, the hydrogel modified membrane consistently maintains a high and stable breakthrough pressure on the level of 6-7 bar (Figure 2b). The breakthrough pressure remains unaltered with variations in SDS concentration, indicating that the breakthrough might be attributed to potential cracks in the hydrogel layer rather than leakage from the membrane's pores. Furthermore, we measured the breakthrough pressure of membrane with different hydrogel thicknesses. The pressure increases with the thickness and achieves 20 bar at the thickness of 750 μm. This remarkable result confirms the excellent anti-wetting performance of the membrane and indicates that breakthrough pressure is primarily determined by the hydrogel layer not the hydrophobic PTFE membrane. Within the hydrogel, water is effectively confined, making it challenging to permeate and wet the underlying hydrophobic PTFE. We have also checked the performance of membranes with various commonly used hydrogels including polyacrylamide (PAAM), poly (hydroxyethyl acrylate) (PHEA) and poly (hydroxyethyl methacrylate) (PHEMA) hydrogels with a gel layer thicknesses of 150 μm–the same as the PSA hydrogel. All of these hydrogels modified membranes show similar high breakthrough pressures, demonstrating the universality of this approach in anti-wetting (Figure 2c).

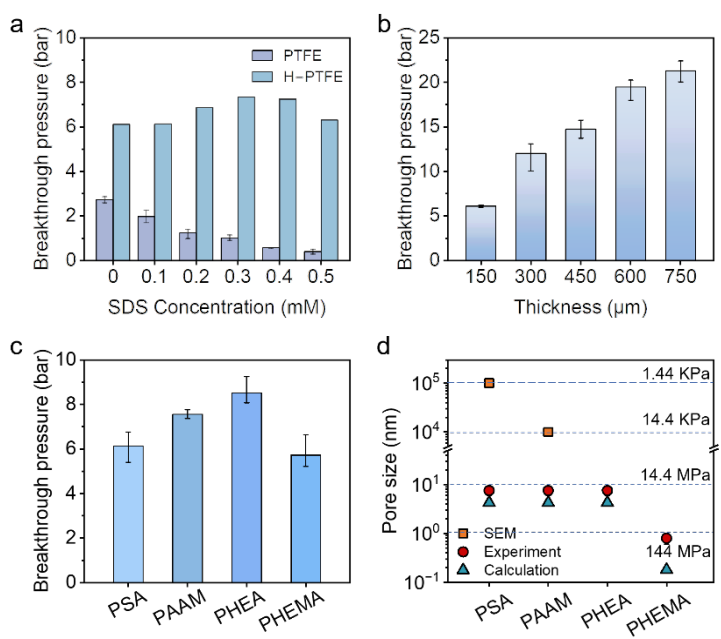

**Figure 2.** Breakthrough pressures of PTFE and H-PTFE membranes. a), Influence of SDS concentration on the breakthrough pressure. Feed solution contains NaCl with a concentration of 0.6 M. b), Breakthrough pressure as a function of the thickness of the PSA layer. c), Breakthrough pressures of different hydrogels including PSA, PAAM, PHEA and PHEMA hydrogel. All of the hydrogels were measured at a thickness of 150 μm. d), Comparison of hydrogel pore radii from SEM observing, diffusion-experiment and theoretical predicting. The blue dotted lines denote the predicted breakthrough pressures at pore radii of $10^0$, $10^1$, $10^4$, $10^5$ nm.

## 2.2. Mechanism of the anti-wetting

Results above highlight two important properties of the H-PTFE membrane: i) high breakthrough pressure; ii) the invariance of the breakthrough pressure with the surfactant concentration. Breakthrough pressure in a porous material can be calculated as[15]

$$P_B = 2\sigma \cos\theta / r \quad (1)$$

where $P_B$ represents the breakthrough pressure, $\sigma$ is the liquid surface tension, $\theta$ is the contact angle of water on the solid material, and $r$ denotes the equivalent hydraulic radius of the pores. In our case, $\theta$ is estimated as zero, given that the hydrogel chains are superhydrophilic (Figure S2, Supporting Information). Hence, the breakthrough pressure is primarily governed by the radius of the pores and the water surface tension. While SEM image (Figure S1b, Supporting Information) of the PSA hydrogel indicates the pore size of ~10 μm, Equation (1) suggests that a pore size of this magnitude cannot yield a high breakthrough pressure as we observed in Figure 2a, c. To accurately determine the true pore size of the hydrogel during its swelling in water, we conducted a diffusion experiment involving particles of varying sizes (Figure S5a, b, and Note S1, Supporting Information). The measured radius of the PSA hydrogel is between 3 and 4.6 nm (Figure S5c, Supporting Information), which indicates a breakthrough pressure of ~10 MPa. This value explains why water is difficult to be squeezed out of the hydrogel. We have also measured the pore size within PAAM, PHEA and PHEMA hydrogels, revealing that all these hydrogels exhibit pore sizes below 5 nm (Figure S5d−f, Supporting Information). These results align closely with theoretical calculations that assume homogeneous distribution of hydrogel chains in water, forming cubic networks (Figure 2d, and Figure S5g−h, Supporting Information). It is thus to say, hydrogels swelling in water, uniformly form pores as small as several nanometers and possess ultra-high breakthrough pressure. The high breakthrough pressure lays the foundation of the anti-wetting characteristics of hydrogel modified hydrophobic membranes. The large pore size observed in SEM images might form during the freeze drying process, and cannot represent the pore structure in the swollen state.[16]

Surface tension of water varies with ions or molecules dispersed inside. Low-surface-tension contaminants naturally accumulate at the liquid-vapor interface and boost the concentration of itself. This surface effect can easily induce membrane wetting. However, the hydrogel modified membrane exhibits an anomalously stable performance even in solutions with high SDS concentrations. To explore the underlying mechanism, we first measured the permeability of SDS within the hydrogel (Note S3, Supporting Information). The SDS transports freely through the hydrogel (Figure S6, Supporting Information). This demonstrates that the hydrogel cannot block the passage of surfactants, allowing surfactants to coexist within the hydrogel during evaporation.

To understand how the surfactant induced wetting is depressed in the hydrogel, we performed DFT calculation and MD simulation to investigate the behavior of surfactants within the hydrogel. The investigated hydrogels include PSA, PAAM, PHEA and PHEMA hydrogels. Two other typical surfactants, hexadecyl trimethyl ammonium bromide (CTAB, cationic surfactant) and sorbitan monolaurate (SPAN 20, neutral surfactant) are also involved. **Figure 3**a shows the binding energies of surfactants with the hydrogel chains and with PTFE. The negative values of binding energies indicate that these interactions are inherently attractive. Almost all the binding energy values for surfactant-hydrogel are larger than those for surfactant-PTFE. Surfactant molecules mainly interact with PTFE molecules via hydrophobic forces, whereas their interactions with the hydrogel polymer chains involve either electrostatic or hydrophobic forces (Figure S7, Supporting Information).

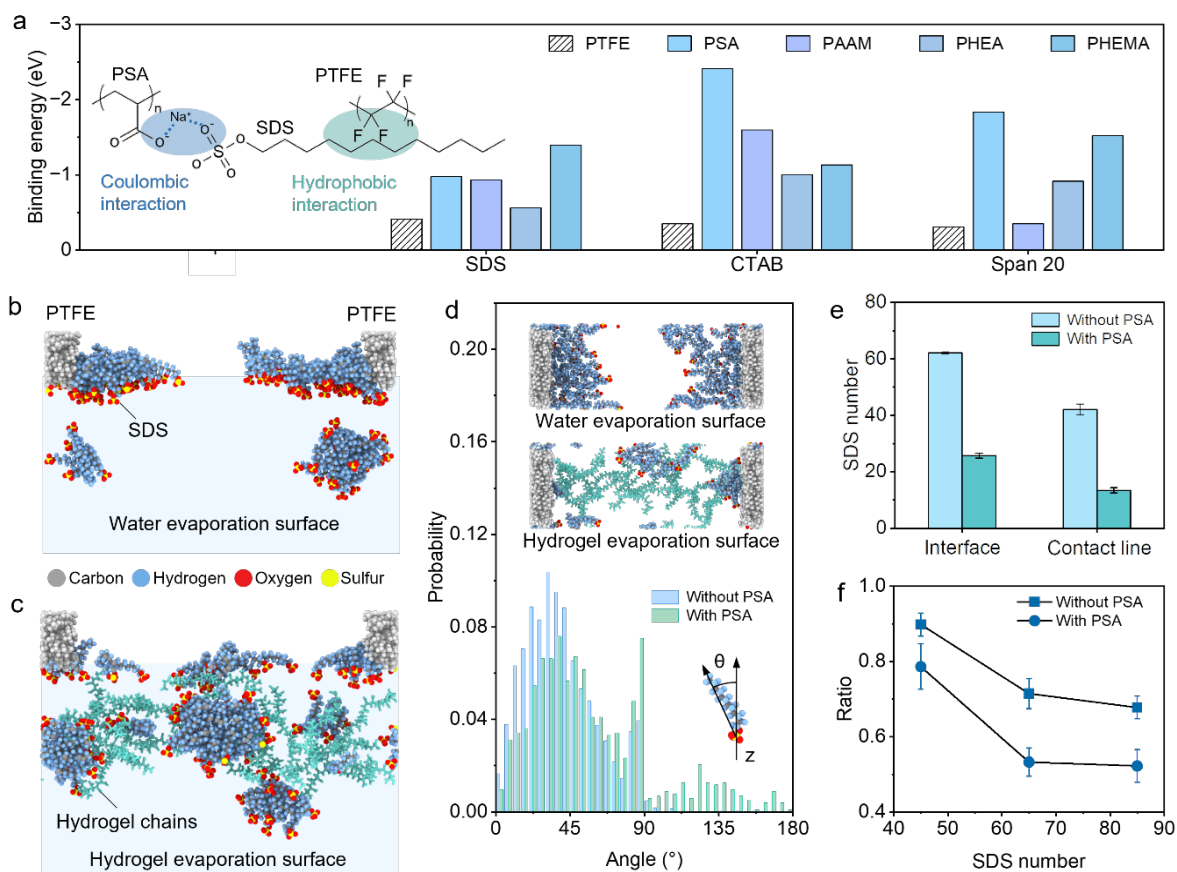

**Figure 3.** Micro-scale anti-wetting mechanism. a), Binding energies between surfactants (SDS, CTAB and SPAN 20) and PTFE or hydrogel chains (PSA, PAAM, PHEA and PHEMA) calculated by density functional theory. Inset figure shows the physical scheme of SDS-PTFE and SDS-PSA molecular interactions. b−c), Side view of distribution state of SDS molecules in PTFE and H-PTFE case, respectively. d), Comparison of SDS orientation angle distributions between PTFE and H-PTFE case. Inset figures are top views of the water evaporation surface (upper) and hydrogel evaporation surface (lower), respectively. e), SDS numbers adsorbed at the liquid-vapor interface and triple-phase contact line. f), Ratio of absorbed SDS number at the triple-phase line to that at the liquid-vapor interface, as a function of the SDS concentration.

The strong interaction between surfactants and hydrogel chains provides a competitive force that hinders the migration of surfactants molecules toward both the liquid-vapor interface and the hydrophobic wall. To characterize the surfactant distribution at the evaporation interface, MD simulations were performed (Figure 3b−f and Figure S8a−b, Supporting Information). Figure 3b depicts the behavior at the liquid-vapor interface without hydrogels, where most of SDS molecules are observed to absorb at the liquid-vapor interface with the hydrophobic end oriented toward the air phase. The small orientation angle (the angle between the sulfur-carbon vector and z-axis, Figure 3d inset) and narrow angle distribution indicate an ordered arrangement (Figure 3d). This surface effect increases the SDS concentration at the liquid-vapor interface and easily lead to wetting.[17] While for PSA hydrogel modified evaporation interface, SDS molecules distribute more uniformly within the liquid layer with fewer molecules accumulating at the evaporation interface (Figure 3c and Figure S8c, Supporting Information). Some SDS clusters are apparently confined within the hydrogel network. The statistical radial distribution function (RDF) between sulfur atoms confirms the uniform distribution of SDS within the PSA (Figure S8d, Supporting Information). The polymer chains rearrange the SDS molecule distribution at the liquid-vapor interface, transitioning it from an ordered to a disordered state. The orientation angle distribution range becomes wider, and even some molecules pull hydrophobic ends into the liquid phase (Figure 3d). We quantified the number of SDS molecules adsorbed at the liquid-vapor interface and three-phase contact line. The presence of hydrogel chains reduces the SDS number at both of these interfaces (Figure 3e). For varying SDS concentrations, the number ratio at the contact line to the whole evaporation interface is apparently lower in hydrogel network compared to pure water (Figure 3f). Hence, it can be concluded that the constraining effect by the hydrogel chains weakens the surfactant accumulation, disrupts the orderly distribution of surfactant at the evaporation surface, and protects the membrane from wetting.

## 2.3. Mechanism of the anti-wetting

To demonstrate potential of the hydrogel modified membrane in practical applications, we conducted membrane distillation using synthetic wastewater feed containing high concentrations of surfactants (144 mg $L^{-1}$), mineral oils (1g $L^{-1}$) and NaCl (192 g $L^{-1}$) (**Figure 4**a inset). As observed in Figure 4a, for commercial PTFE membrane, the flux first raises due to the surfactant induced wetting, then drops due to the oil fouling. The membrane fails after just about 3 minutes of operation. In contrast, the hydrogel modified membrane maintains ultra-stable performance. The surface of the membrane keeps clean even after 96 hours of continuous distillation (Figure 4b). Oil fouling is negligible due to the oleophobicity of the hydrogel (Figure 4c). The treated water has the electric conductivity of 1.68 μS $cm^{-1}$, two orders of magnitude smaller than that of the initial wastewater (209 mS $cm^{-1}$) (Figure 4d, e). The concentration of $Na^+$ in the distilled water meets the drinking water standards prescribed by the World Health Organization (WHO).[18] Raman test shows no signal for C-H bond and thus indicates no organic pollutes in the distilled water (Figure 4f). All these results demonstrate high quality of the treated water and underscore the stable performance of the membrane, manifesting potential application of the membrane in hypersaline wastewater treatment towards zero liquid discharge.

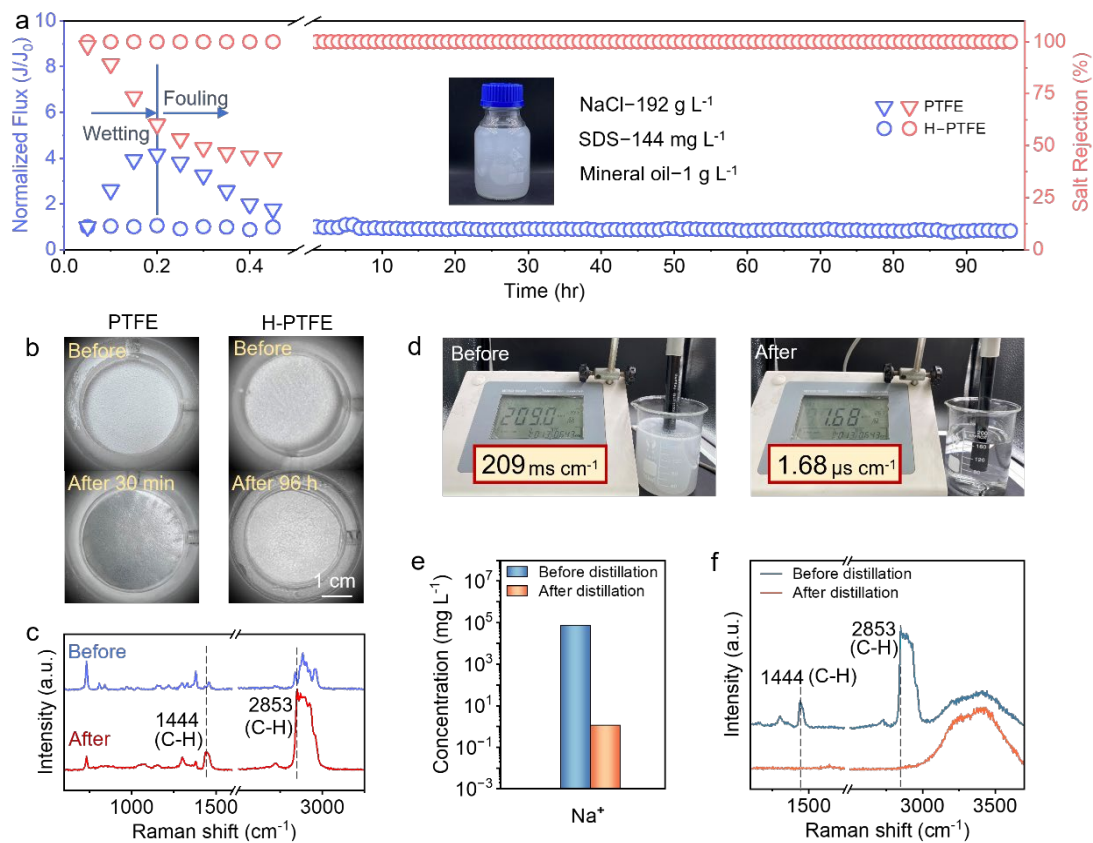

**Figure 4.** Micro-scale anti-wetting mechanism. Demonstration of multi-component wastewater distillation with the H-PTFE membrane. a), Normalized vapor flux (blue points) and salt rejection rate (red points) of the PTFE and H-PTFE membrane in DCMD operation using feed containing 192 g L$^{-1}$ NaCl, 144 mg L$^{-1}$ SDS, and 1 g L$^{-1}$ mineral oil. The feed and permeate temperatures were set as 60 ºC and 20 ºC, respectively. The initial vapor flux of the H-PTFE membrane experiment was 10.33 L m$^{-2}$ h$^{-1}$. b), Photos of permeate sides of PTFE and H-PTFE membrane before and after distillation, respectively. c), Raman spectra of PTFE membrane at the permeate side before and after distillation. The Raman shift peaks at 1444 cm$^{-1}$ and 2853 cm$^{-1}$ correspond to the symmetric stretching and asymmetric deformation of C-H bond. Existence of C-H bond indicates the oil pollution of the membrane surface. d–f), Conductivities, Na$^+$ concentrations and Raman spectra of solution before and after distillation, respectively.

## 3. Conclusion

In summary, we have proposed a novel strategy to mitigate the surface accumulation and redistribute low-surface-tension contaminants at the evaporation interface, achieved simply by incorporating a hydrogel layer. The as-fabricated composite membrane shows stable performance in the presence of a high salt concentration of 5 M and surfactant concentration of 8 mM. The breakthrough pressure of the membrane is as high as 20 bars with the existence of surfactants. DFT calculation and MD simulation unveil that the interaction between hydrogel chains and contaminants mitigates the surfactant accumulation, disrupts the orderly distribution of surfactant at the evaporation surface, and protects the membrane from wetting. We also demonstrate stable performance of the membrane's practical utility in processing synthetic wastewater containing high concentration of surfactant, mineral oils, and NaCl, showing its potential applicability in hypersaline water treatment and zero-liquid-discharge.

## 4. Experimental Section/Methods

*Chemicals and Materials*: In fabrication of PSA, PAAM and PHEA hydrogel, sodium acrylate (SA), acrylamide (AAm), and hydroxyethyl acrylate (HEA) were used as reaction monomers respectively. N, N′-methylenebis (acrylamide) (MBAA) was used as the crosslinking agent. Tetramethylethylenediamine (TEMED) was used as the catalyst. Ammonium persulfate (APS) was used as the thermal initiator. Preparation of PHEMA hydrogel is the same as previous literature.[19] Commercial PTFE membranes were purchased from Sterlitech Co., Ltd. Gold nanoparticles were purchased from Qingdao New Aurora Future Hi-Tech Co., Ltd. Sodium dodecyl sulfate (SDS), mineral oil, bovine serum albumin, α-chymotrypsin, D-tryptophan, carbamide, and other chemicals used in hydrogel fabrication were purchased from Sigma-Aldrich. Deionized water (18.3 MΩ) was used in all experiments. In addition, include a section titled Statistical Analysis at the end that fully describes the statistical methods with enough detail to enable a knowledgeable reader with access to the original data to verify the results. The values for N, P, and the specific statistical test performed for each experiment should be included in the appropriate figure legend or main text.

*Fabrication of the H-PTFE membrane*: The H-PTFE membrane was fabricated through graft copolymerization between hydrogel and the polypropylene layer of the commercial PTFE membrane (Figure S1a, Supporting Information).[20] Specially, for a PSA hydrogel, the hydrogel precursor solution containing 2 mol L$^{-1}$ monomer SA, 0.08 mol L$^{-1}$ MBAA and 0.008 mol L$^{-1}$ TEMED was first vacuumed to remove oxygen, then 0.008 mol L$^{-1}$ APS was added to the solution. At the same time, the polypropylene side of the PTFE membrane was treated with oxygen plasma at a power of 18 w for 5 minutes to create active functional groups on the polypropylene fibers. Finally, the mixed pre-solution was dropped on the plasma-treated substrate and synthesized in N$^2$ environment for 4 h. The thickness and size of the fabricated hydrogel were controlled through a standard mold. The gas pressure and temperature in polymerization were controlled at 0.06 MPa and 30 °C, respectively. H-PTFE membranes of PAAM and PHEA were prepared in the same way as the PSA described above.

*Characterizations*: The surface and cross-sectional morphologies of the hydrogels and H-PTFE membranes were characterized by the scanning electron microscopy (SEM, MIRA 3, TESCAN). Components and elements were characterized by the energy dispersive spectroscopy (EDS, Aztec Energy, Oxford Instruments). Copolymerization between the hydrogel and PTFE was verified by the Fourier transform infrared spectroscopy (FTIR, FTIR 5700, Thermo Co., Ltd.). Static contact angle tests were performed using an optical tensiometer (OCA25, Dataphysics). Conductivity of the permeate in membrane distillation was monitored in real time by a flow-through conductivity meter (ET908, eDAQ). Absorbance of protein molecules and gold nanoparticles were obtained by the UV-visible spectrophotometer (LAMBDA 1050+, PerkinElmer). Ionic conductivities of solutions in static experiments were measured using a conductivity meter (Mettler Toledo, FE38). Ion concentration was measured using an atomic absorption spectrometer (contrAA700, Analytikjena). Organic foulants were characterized by a laser confocal Raman spectrometer (Alpha 300 RA, Oxford).

*Evaluation of membrane distillation performance*: DCMD experiments were used to evaluate the performance of the PTFE and H-PTFE membrane. The schematic of the experimental setup is shown in Figure S3. In all DCMD experiments, the temperatures of the feed and distillate were maintained to be 60 °C and 20 °C, respectively. Flow rates of the feed and distillate solutions were controlled at 0.3 and 0.15 L min$^{-1}$. The membrane was fixed in the middle of a module with an effective mass transfer area of 7.07 cm$^2$ (diameter of 30 mm). Mass and conductivity variations of the permeate were recorded in real time using an electronic balance and conductivity meter, respectively. The inlet pressure was controlled using a digital pressure sensor. In DCMD experiments, the hydrogel layer contacted with the feed. The feeds in wetting and fouling tests were prepared as follows: NaCl was firstly dissolved in distilled water, then the organic foulants were added to the solutions and mixed under ultrasonic crushing for 1 hour using a cell crusher.

*Breakthrough Pressure Measurement*: The measured membrane was sandwiched between two stainless-steel cells. The feed cell was filled with solution and connected with a compressed nitrogen cylinder. The permeate cell was exposed to air (Figure S4a, b, Supporting Information). A high-resolution microscopic camera was used to monitor the breakthrough liquid (Figure S4c, Supporting Information). During the measurements, an initial pressure of 0.5 bar was applied to the membrane and held for 5 min to obtain a uniform pressure. Then, the pressure was increased at an interval of 1 bar. At each pressure interval, the membrane was kept at a constant pressure for 5 mins. When the first water droplet was observed at the permeate side, the corresponding pressure was considered to be the breakthrough pressure.

*Density functional theory calculation*: All binding energies between polymer chains and surfactants were calculated through VASP code. PBE-GGA method was adopted to calculate the electronic exchange-correlation energy, while the ionic cores were described by PAW method. DFT-D3 method was employed to describe the weak interaction. Kinetic energy cut-off was specified as 450 eV. The convergence criterion for the electronic structure iteration was set to be $10^{-5}$ eV, and that for geometry optimizations was set to be 0.05 eV Å$^{-1}$. A Gaussian smearing of 0.1 eV was applied for geometry optimization and energy computations. For all models, the cell size is set as 14.75×12.78×22.04 Å$^3$, with a k-point mesh of 2×2×1. All the five polymer chains periodically extend in x-axis direction. Each chain contains 6 monomers in the cell. The binding energy between surfactants and PTFE was calculated in gas-phase environment, and the binding energy was defined as $E_b = E_{tot} - E_{sur} - E_{PTFE}$. While the binding energy between surfactants and hydrogel chains was calculated with explicit water model. We controlled the density of the cell to be about 1.0 g cm$^{-3}$ through adjusting the number of water molecules in the cell. Considering the interaction between hydrogel chains and surfactants within explicit water model, the binding energy was defined as $E_b = E_{tot} - E_{gel-w} - E_{sur-w} + E_w$.[21]

*Molecular Dynamics Simulations*: We constructed two quasi-2D models similar to previous work.[22] One model consists of PTFE walls, Na$^+$ ions, Cl$^-$ ions, SDS and water molecules, and the other model adds PSA hydrogel on the basis of the former (Fig. S8a, b, Supporting Information). In these models, there are both gas-liquid interfaces and solid-liquid-gas three-phase contact lines. According to the method proposed by Velioglu et al,[23] the two 1×4×5 nm$^3$ PTFE walls were built through packmol package,[24] each wall contained 265 PTFE monomers, the resulting density of 2.2 g cm$^{-3}$ is in consistent with experimental value. Both the two systems were initially filled with 85 SDS molecules to simulate the real state at the gas-liquid interface when the concentration reaches the critical micelle concentration of SDS.[9b] We also varied the number of SDS to investigate the distribution of SDS at the interface and three-phase contact line under different SDS concentrations. Total number of water molecules was 9832 and the concentration of NaCl was set as 4.5 M (in consistent with experiment). All the molecular dynamic simulations were performed through LAMMSP package.[25] We adopted periodic boundary in x-axis and y-axis directions. The water model was specified as SPC/E, while all the other atoms were described by OPLS-AA force field.[26]

The cut-off distance for pair interaction was set as 1.2 nm. Long range columbic force was calculated through PPPM method. NVT ensemble with a timestep of 1 fs was employed in all simulations, and the temperature was set as 300 K. After the equilibrium calculation of 10 ns, an additional 1 ns was performed to generate production data. The SDS orientation angle is defined as the included angle between z-axis positive direction and the straight line formed by S (sulfonate) and C (tail carbon in dodecyl) atoms. The visualization of molecular model was conducted through VMD and OVITO.[27]

# Supporting Information

**Hydrogel modified evaporation interface for highly stable membrane distillation**

*Yanni Ma†, Zehua Yu†, Xifan Fu, Zhi Huang\*, Tenghui Qiu, Na Zhao, Huidong Liu, Kang Liu\**

**Contents**

**Note S1. Measurement of the effective pore size in hydrogels.**

The pore size range was measured by screening particles with different sizes to diffuse through the hydrogel.[28] Test hydrogel layers with the effective area of 3.14 cm$^2$ were assembled with two chambers (Figure S5a). The two chambers were filled with solutions (feed side) and DI water (diffusion side), respectively. Specially, particles or molecules with fixed radius were dispersed into the feed solution. A UV-VIS spectrometer was used to detect the particles or molecules after 48-hour stirring (Figure S5a, b, Supporting Information). The pore size range was determined by screening the particles or molecules with different sizes (Figure S5c−f, Supporting Information). In the tests, we employed gold nanoparticles (7.5 nm), bovine serum albumin (4.6 nm), α- chymotrypsin (3 nm), tryptophan (0.6 nm) and carbamide (0.3 nm) to provide a test range from 0.3 to 7.5 nm.[29]

**Note S2. Hydrogel pore-size prediction.**

The hydrogel pore size was theoretically predicted assuming the hydrogel chains are homogeneously distributed in water and form cubic networks. The polymer chain was assumed to be rigid and in a cylinder shape with a bottom diameter d. A cube unit with a side length of a consists of three hydrogel polymer chains and a cross-linker molecule (Figure S5f, Supporting Information). In this unit, the polymer chain length is equivalent to a. The cross-linker can be neglected since its volume proportion is much lower than the polymer chains. The volume fraction of the hydrogel chain, $\phi$, can be expressed as $\phi = (3/4 \pi d^2 a) / a^3$. $\phi$ can also be expressed as $\phi = m / \rho V$, where m is the monomer mass, $\rho$ is the monomer density and V is the total volume of the precursor solution. By combining the two equations, we obtain $a = (3 \pi \rho V / 4 m)^{1/2} d$. In order to obtain d, molecular dynamic simulations were conducted to obtain the statistical radial distribution function (RDF) between the oxygen atoms in water molecules and the carbon atoms in skeletons of hydrogel polymer chains (Figure S5g, Supporting Information). The distance of the first valley from the origin point was considered to be the bottom radius of the polymer chain.

**Note S3. SDS diffusion measurement**

Diffusion experiments were conducted to determine whether the SDS can permeate through the PSA hydrogel (Figure S6, Supporting Information). The hydrogel membrane with an area of 3.14 cm$^2$ was assembled with two chambers. The two chambers were filled with 5 mM SDS aqueous solution (feed side) and DI water (diffusion side), respectively. The electric conductivity of solution at the diffusion side was recorded at an interval of 15 mins.

**Figure S1. Measurement Preparation and characterization of the H-PTFE membrane.**

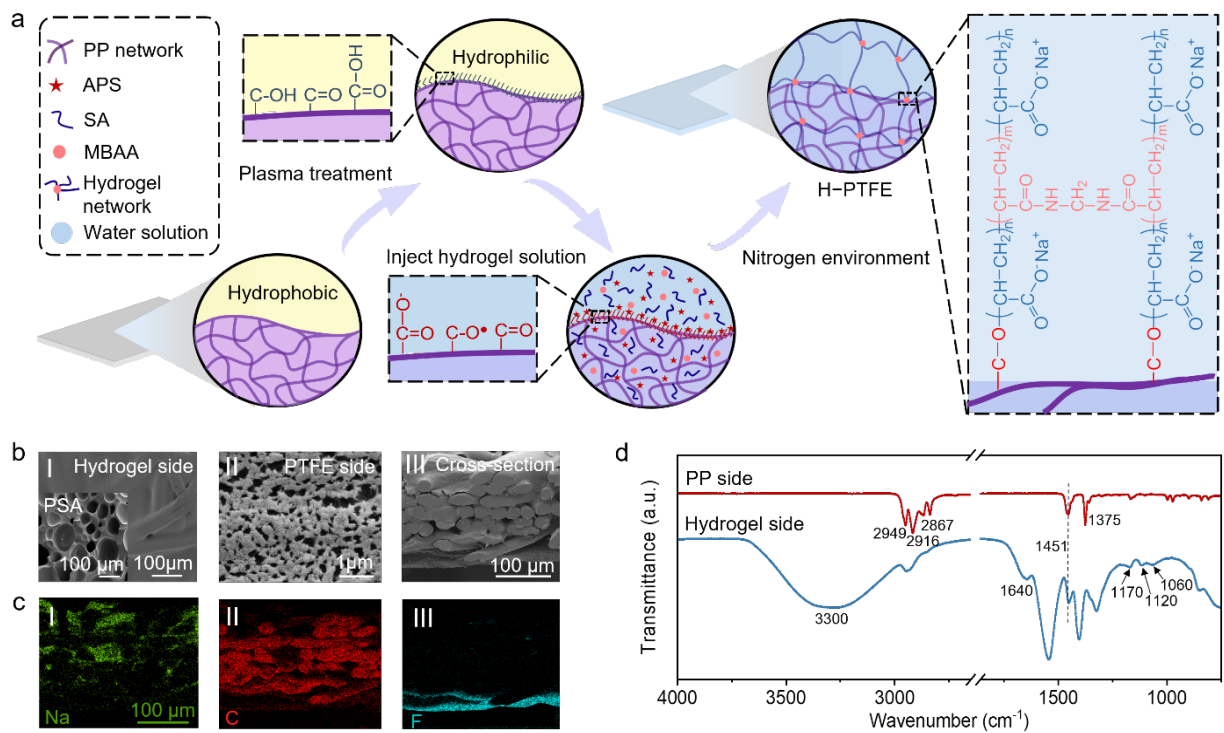

**Figure S1.** Measurement Preparation and characterization of the H-PTFE membrane. a), Fabrication process of the H-PTFE membrane. b), SEM images of the hydrogel side (I), PTFE side (II) and cross-sectional view of the H-PTFE (III). Inset in b(I) is the SEM image of the freeze-dried PSA hydrogel. c), EDS of the cross-sectional view of the H-PTFE membrane. Na (I), C (II) and F (III) are the characterized elements of the PSA, PP and PTFE, respectively. d), ATR-FTIR spectra of the PP side (red curve) and the hydrogel side of the H-PTFE (blue curve) membrane.

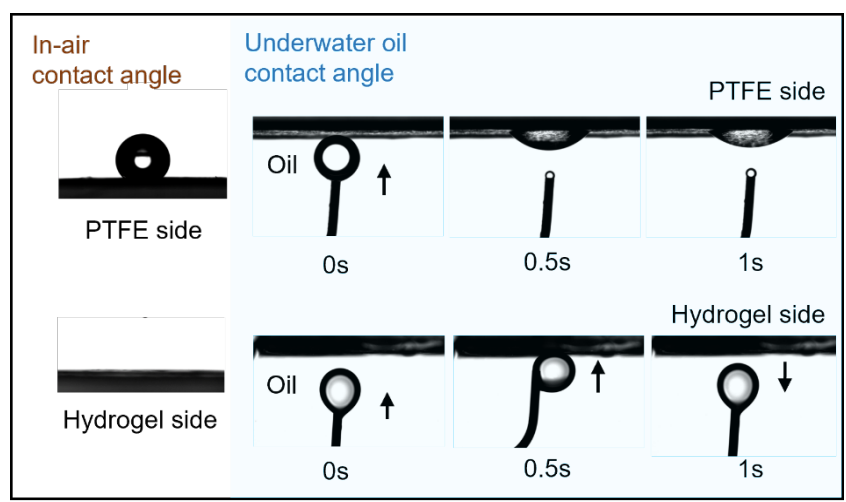

**Figure S2.** Water and oil contact angles of the H-PTFE membrane in-air and underwater.

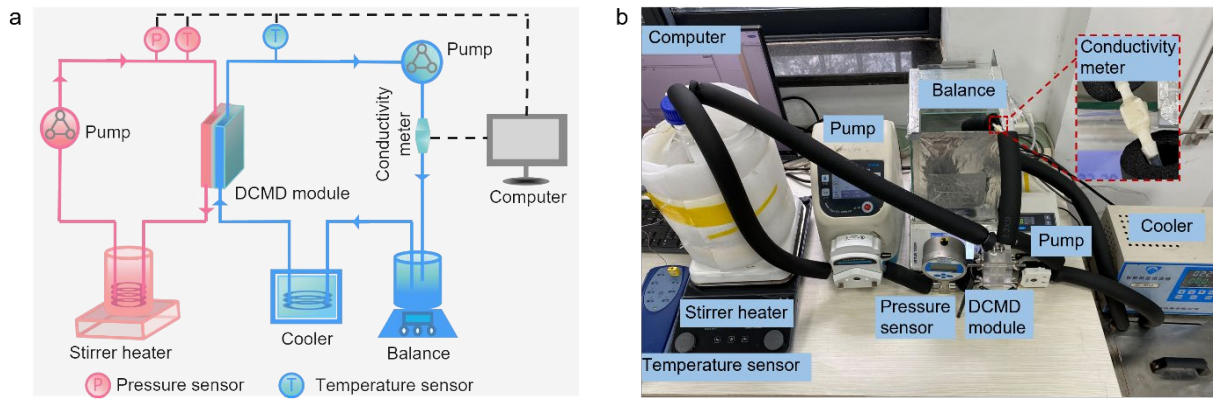

**Figure S3.** Membrane distillation and breakthrough pressure measurement. a), Schematic of the setup of direct contact membrane distillation (DCMD). b), Photograph of the DCMD measuring system.

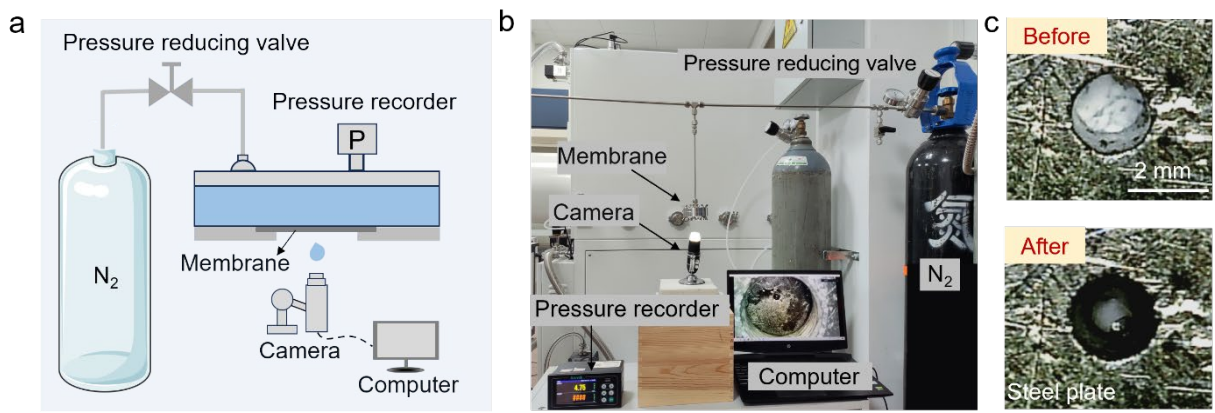

**Figure S4.** Breakthrough pressure measurement. a), Schematic illustration of the breakthrough pressure measuring system. b), Photograph of the breakthrough pressure measuring system. c), Photographs of the PTFE before and after breakthrough pressure test.

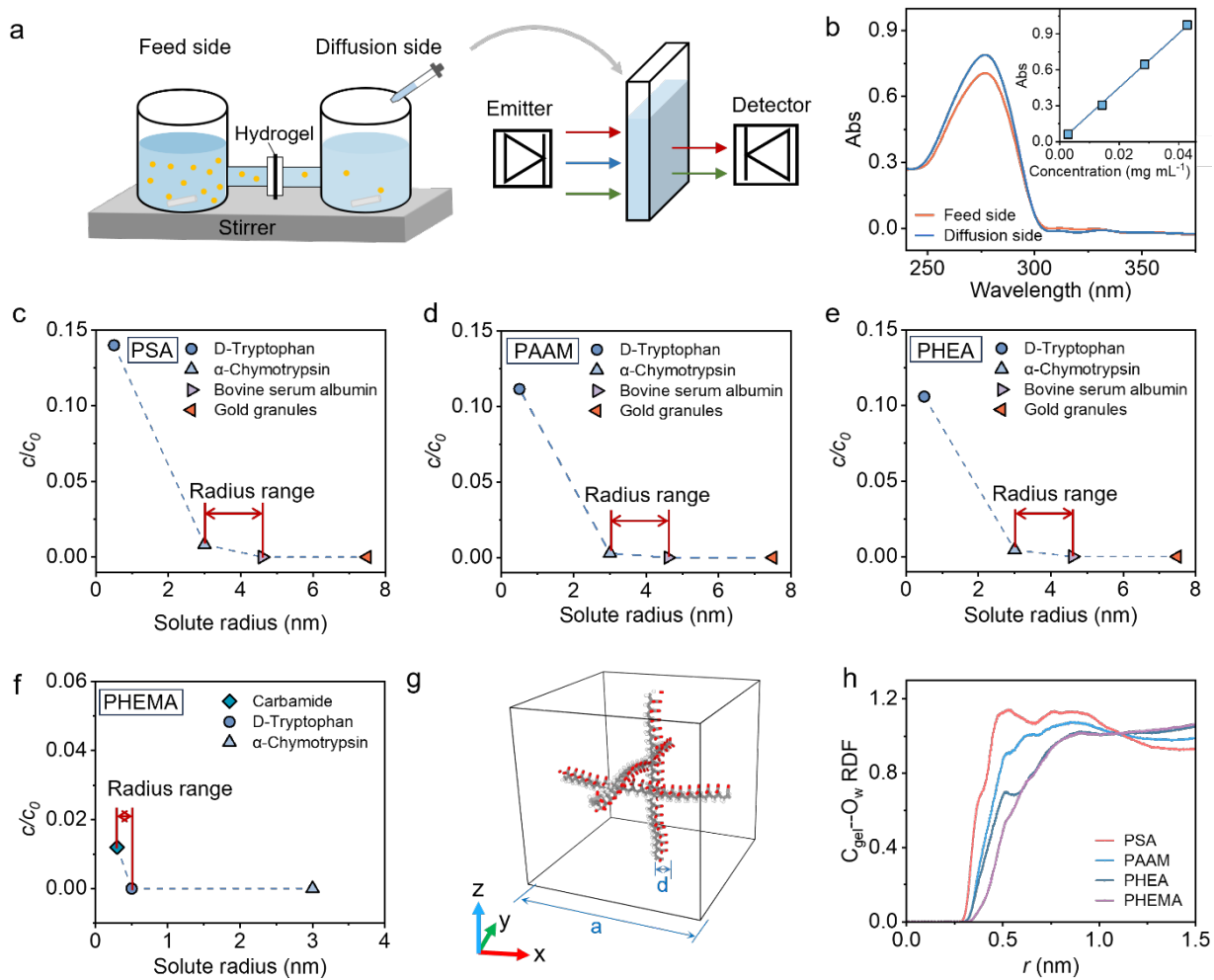

**Figure S5.** Determination of the pore size of the hydrogel. a), Experimental setup of the diffusion-screening method to measure the pore size of the hydrogel. b), Absorbances of solution s at the feed side (6x dilution) and diffusion side (35x dilution) after 48h stirring. Inset figure shows the calibration between the absorbance and concentration. c-f), Measured radius ranges of PSA, PAAM, PHEA and PHEMA. $c/c_0$ represents the ratio of the molecular concentration of diffusion side after 48h stirring to the initial concentration of the feed side. g), Diagram of a cubic hydrogel network under swelling state. h), RDFs between the oxygen atoms in water molecules and the carbon atoms in skeletons of hydrogel polymer chains.

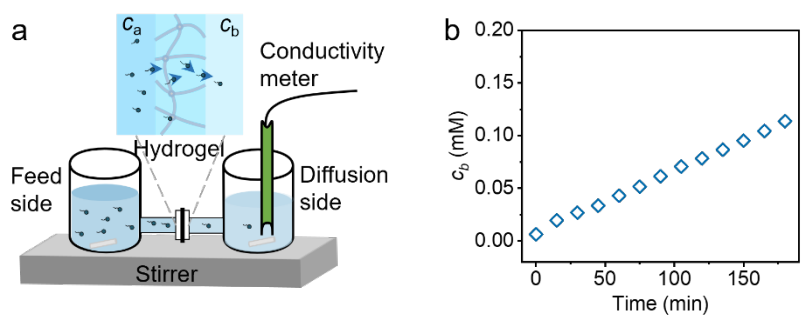
21

**Figure S6.** Determination of the pore size of the hydrogel. a), Experimental set-up to measure the diffusion ability of the SDS in hydrogel. ca and cb represent the SDS concentration in the feed side and diffusion side, respectively. b), SDS concentration cb in the diffusion side as a function of diffusion time.

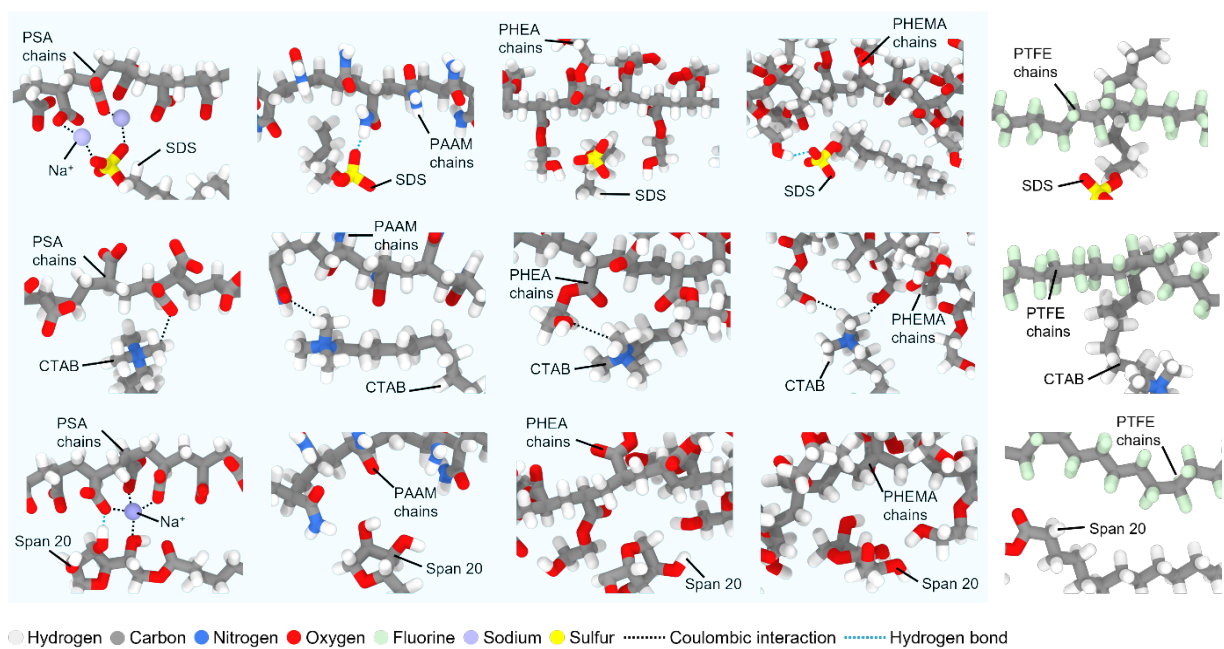

**Figure S7.** Diagrams of interactions between surfactants and hydrogel chains or PTFE chains. The hydrophobic end of the surfactant interacts with PTFE chains through hydrophobic interaction. The geometry of the two shows vertical (SDS, CTAB) or parallel (Span 20) arrangement. Hydrophobic interaction also exists between surfactants and the hydrogel skeleton (e.g., PHEA-SDS, PAAM-Span 20, PHEA-Span 20, PHEMA-Span 20). Since the surfactant is dispersed in the hydrogel network, there are more effective interaction area and more complex interaction morphology. Meanwhile, the hydrophilic end of the surfactant interacts with the polar functional groups of the hydrogels through coulombic interaction (PSA-SDS, PSA-CTAB, PAAM-CTAB, PHEA-CTAB, PHEMA-CTAB, PSA-Span 20) or hydrogen bonds (PAAM-SDS, PHEMA-SDS, PSA-Span 20).

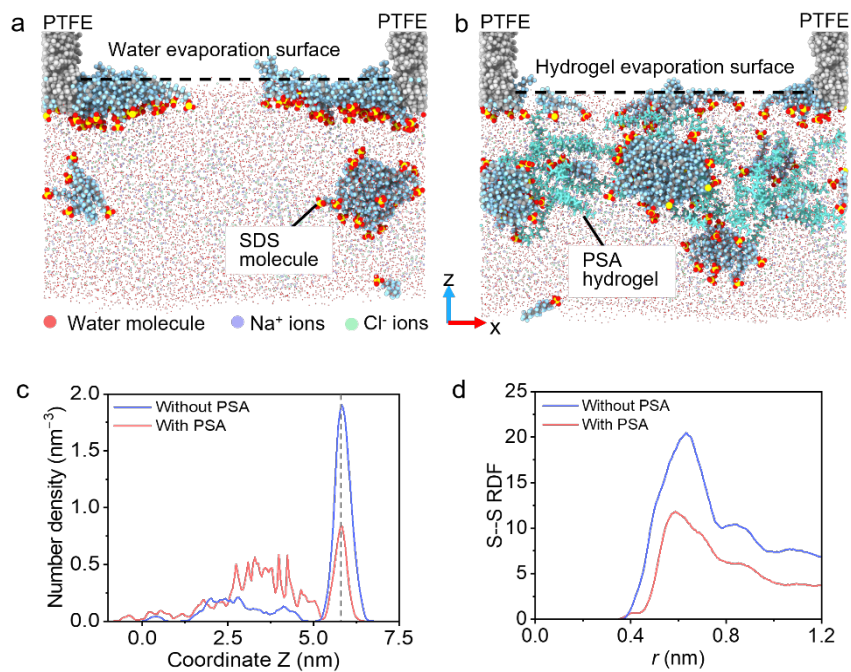

**Figure S8.** Distribution states of SDS near the evaporation surface. a−b), Side view of the distribution state of SDS molecules with and without PSA, respectively. c), Number density of SDS along z-axis direction. d), Statistical radial distribution functions (RDFs) between sulfur atoms in SDS molecules.

| Membrane | Type | Feed solution | | Year | Ref. |
|---|---|---|---|---|---|
| | | NaCl | SDS | | |
| SiNPs-CTS/PFO/SiNPs/CTAB/PVDF-HFP | JM | 0.6 M | 0.4 mM | 2017 | [23] |
| Catechol/Chitosan/PVDF | JM | 3.5 wt% | 50 mg L$^{-1}$ | 2021 | [24] |
| PVA/PVDF | JM | 35 g L$^{-1}$ | 0.2 mM | 2021 | [25] |
| PVA/Al$_2$O$_3$ NPs/PVDF | JM | 35 g L$^{-1}$ | 0.1 mM | 2022 | [26] |
| Teflon®AF1600/PDA/PTFE/PP | JM | 3.5% | 0.6 mM | 2020 | [27] |
| PVA/TA/PTFE | JM | 3.5 wt% | 0.4 mM | 2023 | [28] |
| 120/15FC4/PVDF | OM | 1 M | 0.5 mM | 2022 | [29] |
| SiNPs/glass fiber membrane | OM | 1 M | 0.4 mM | 2014 | [30] |
| nano-ZnO needles/PVDF | OM | 50 g/L | 0.3 mM | 2022 | [31] |
| PSA hydrogel/PTFE/PP | H-PTFE | 5 M | 8 mM | | This work |

**Table S1.** Distribution states of SDS near the evaporation surface. Comparison of feed salt and SDS concentration of membrane distillation with the literature data working with omniphobic membranes (OMs) and Janus membranes (JMs).

Supplementary References